\documentclass[twocolumn,a4paper,showpacs,superscriptaddress,aps,prl]{revtex4-1}
\usepackage{amsmath}
\usepackage{graphicx}
\usepackage[english]{babel}
\usepackage{subfigure}
\usepackage{calc}
\begin{document}

\title{Low energy graphene edge termination via small diameter nanotube
formation}

\author{Viktoria V. Ivanovskaya}
\email{v.ivanovskaya@gmail.com}
\affiliation{Institut des Mat\'{e}riaux  Jean Rouxel (IMN), UMR 6502 CNRS,
University of Nantes, 44322 Nantes, France}
\affiliation{Institute of Solid State Chemistry, Ural division of Russian
Academy of Science, 620041, Ekaterinburg, Russia}

\author{Alberto Zobelli}
\affiliation{Laboratoire de Physique des Solides, Univ. Paris-Sud, CNRS UMR
8502, F-91405, Orsay, France}

\author{Philipp Wagner}
\affiliation{Institut des Mat\'{e}riaux  Jean Rouxel (IMN), UMR 6502 CNRS,
University
of Nantes, 44322 Nantes, France}

\author{Malcolm I. Heggie}
\affiliation{Department of Chemistry, University of Sussex, Falmer, Brighton
BN1 9QJ, United Kingdom}

\author{Patrick R. Briddon}
\affiliation{School of Electrical, Electronic and Computer Engineering,
University of Newcastle upon Tyne, Newcastle NE1 7RU, United Kingdom}

\author{Mark J. Rayson}
\affiliation{Dept. Eng. Sciences and Mathematics, Lule\r{a} University of Technology, S-97187 Lule\r{a}, Sweden}

\author{Chris P. Ewels}
\email{chris.ewels@cnrs-imn.fr}
\affiliation{Institut des Mat\'{e}riaux  Jean Rouxel (IMN), UMR 6502 CNRS,
University
of Nantes, 44322 Nantes, France}

\begin{abstract}
We demonstrate that free graphene sheet edges can curl back on themselves, 
reconstructing as nanotubes.  This results in lower formation energies
than any other non-functionalised edge structure reported to date in the
literature. We determine the critical tube size and formation barrier and
compare with density functional simulations of other edge terminations including
a new reconstructed Klein edge.
Simulated high resolution electron microscopy images show why such rolled edges
may be difficult to detect. Rolled zigzag edges serve as metallic conduction
channels, separated from the neighbouring bulk graphene by a chain of insulating
sp$^3$-carbon atoms, and introduce Van Hove singularities into the graphene
density of states.  

\end{abstract}

\maketitle

The atomic structure of graphene and graphene edges, is a subject
of great interest, in particular stimulated by new aberration corrected atomic
resolution electron microscopy studies \cite{RevEdges,girit2009graphene}. 
Edge structure can
define the chemical and electronic properties of graphene ribbons 
\cite{enoki2007electronic, *kunstmann2011stability, *wassmann2008structure},
yet there is no consensus about the most stable free edge
structure. 
Unterminated edges, consisting in a line of atoms with dangling bonds, are
inherently unstable and subject to chemical functionalisation in ambient.  However
under vacuum conditions, such as in electron microscope columns, unterminated
edges can be observed.  Simply cutting through the graphene lattice results in
the most studied edge structures: armchair or zigzag,
or a combination of the two.
The first way that free edges can stabilise themselves is via rehybridisation
of the carbon atoms.  This occurs spontaneously for the armchair edge giving a sequence
of double bonds along the edge.  The
zigzag edge has been shown to be also metastable and undergo a 5-7
reconstruction \cite{Pekka}, which has been recently experimentally
confirmed \cite{koskinen2009evidence,*chuvilin2009graphene}. 
High resolution transmission electron microscopy (HRTEM) suggests that
alternative edge structures may be
also common, including
the theoretically less stable zigzag and Klein edges \cite{PhysRevLett-Kazu,
Warner, Huang, girit2009graphene, gass2008free, meyer2007structure,
rotkin2002analysis}.  Other reconstructed edges, 
loosely based on the Haeckelite
structures \cite{terrones2000new} have also
been proposed \cite{RevEdges}.
Besides reconstruction, HRTEM \cite{PhysRevLett-Kazu, Warner-2, Warner, Huang,
girit2009graphene, gass2008free, meyer2007structure, rotkin2002analysis,
meyer2007roughness} has shown that  free edges can fold back on themselves
resulting in  no longer a graphene monolayer (so-called grafold
\cite{grafold}).
The energetic cost of bending the layer is partially compensated by Van der
Waals interactions in the stacked region.

In this Letter we extend the range of edge types, introducing a third
type of stabilised edge
whereby edges are rolled back on themselves and rebonded into the graphene
sheet.  This results in a graphene
monolayer  with  nanotube at the edge site, and
eliminates all dangling bonds by
sp$^3$-like rehybridisation of the carbon atoms along the rebonding line.

We perform spin polarized density functional calculations under the local
density approximation using the AIMPRO code
\cite{Aimpro,*Aimp-2, *Aimp-3}.
Graphene edges are modeled using ribbons with width $\sim$50\AA{}
in orthogonal supercells large enough to avoid interaction between neighbouring
cells.
Tubular rolled edges were created symmetrically on both of the two ends of
the graphene ribbon.
Edge formation energies per unit length (eV/\AA) are determined via
\begin{equation*}
E_{form} = \frac{E_{tot}-n\cdot \mu_C}{2L}
\label{formenergy}
\end{equation*}
where E$_{tot}$ is the total internal energy of a system with $n$
carbon atoms, $\mu_C$ is the energy of a carbon atom in a perfect
graphene sheet and $L$ is the ribbon edge length.
Reaction barriers were calculated using the climbing nudged elastic band
technique \cite{neb-1}, with all atoms allowed to move. More details are given in the supplementary materials.

\begin{figure}[tbp]
\includegraphics[angle=270, width=\columnwidth]{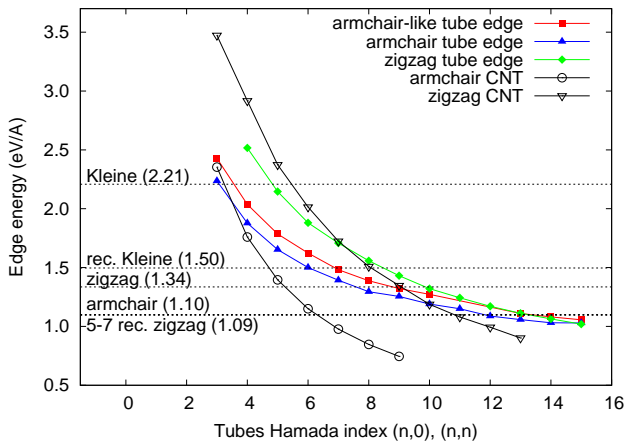}
\caption{(Color online) Formation energy (eV/\AA) for different graphene
edges (dashed
lines indicated on the figure: Klein, reconstructed Klein, zigzag,
armchair and 5-7 reconstructed zigzag),  free standing zigzag
 and armchair  tubes, armchair-,
armchair-like-, and zigzag-tube terminated edges.}
\label{formenerg}
\end{figure}

We begin by considering formation energies for zigzag,
armchair, 5-7 reconstructed zigzag, Klein\cite{Klein} and reconstructed Klein
edges (Fig. \ref{formenerg} and supplementary materials).
In all cases no out of plane distortion (e.g. edge rippling) was 
found after relaxation.
In agreement with results of Ref. \cite{Pekka},   we obtain
the armchair and 5-7 reconstructed zigzag edges to be the most energetically
stable. The least stable is the Klein edge with unsaturated carbon
atoms, however if symmetry is allowed to break, it
spontaneously reconstructs by rebonding in pairs, gaining
$\sim$0.7 eV/{\AA}  (see supplementary materials).
Segments of unreconstructed Klein edge were recently observed by annular dark
field microscopy \cite{PhysRevLett-Kazu, suenaga2010atom} but the observed Klein
edges might have residual hydrogen not detectable by annular
dark field microscopy.

\begin{figure}
 \subfigure{\frame{
 \includegraphics[width=0.3\linewidth]{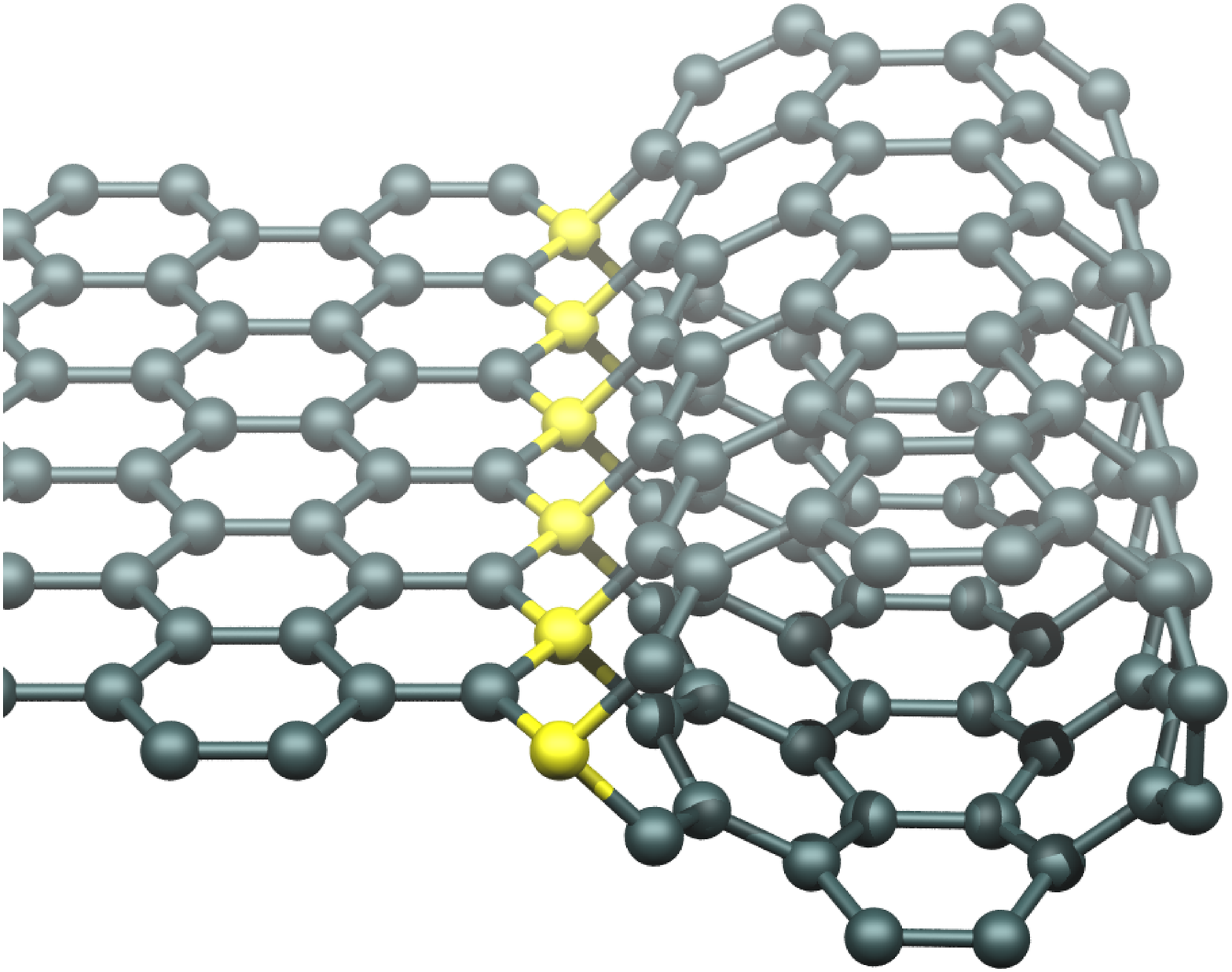}
}
 }
  \subfigure{
  \frame{
 \includegraphics[width=0.3\linewidth]{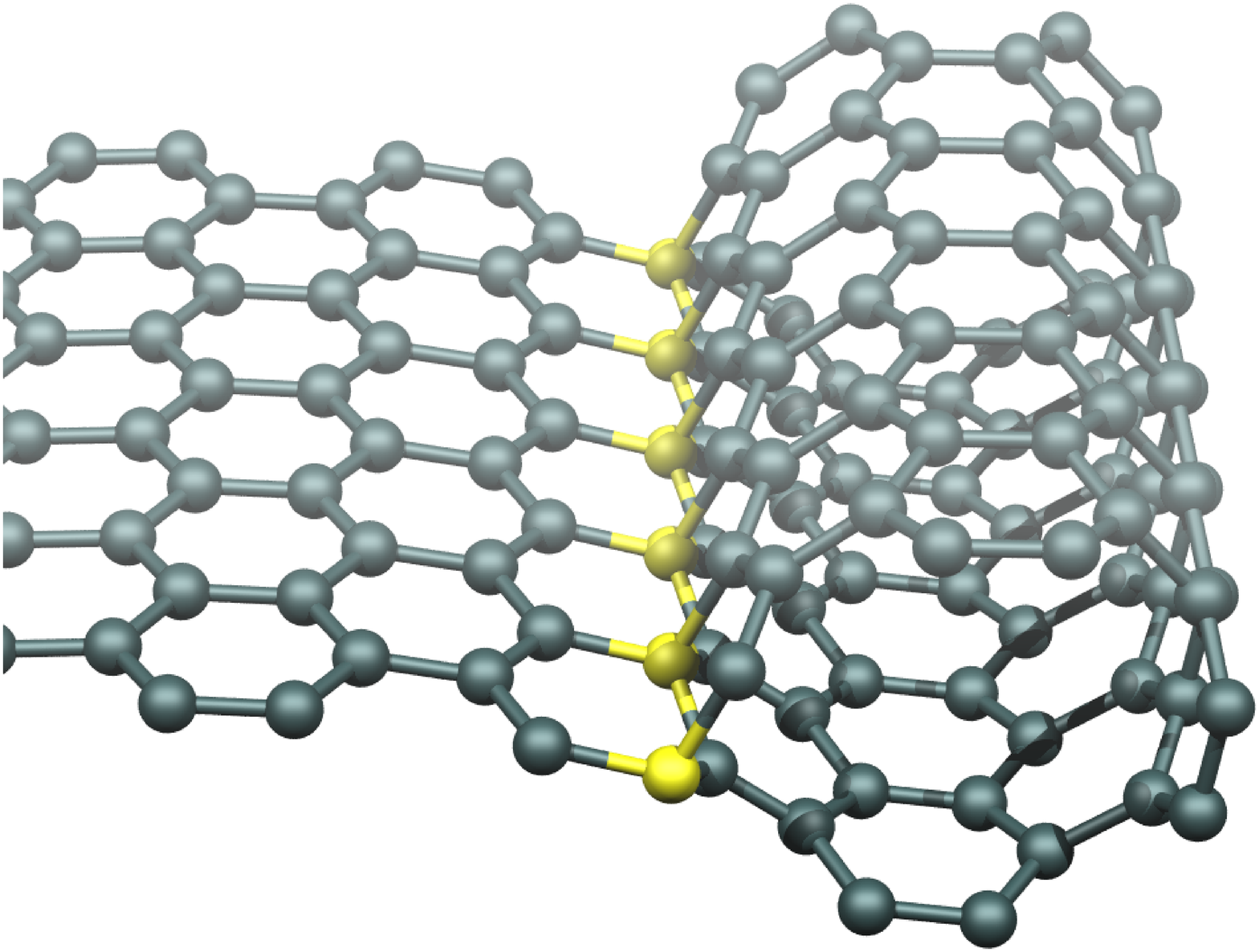}
 }
 }
\subfigure{\frame{\includegraphics[width=0.3\linewidth]{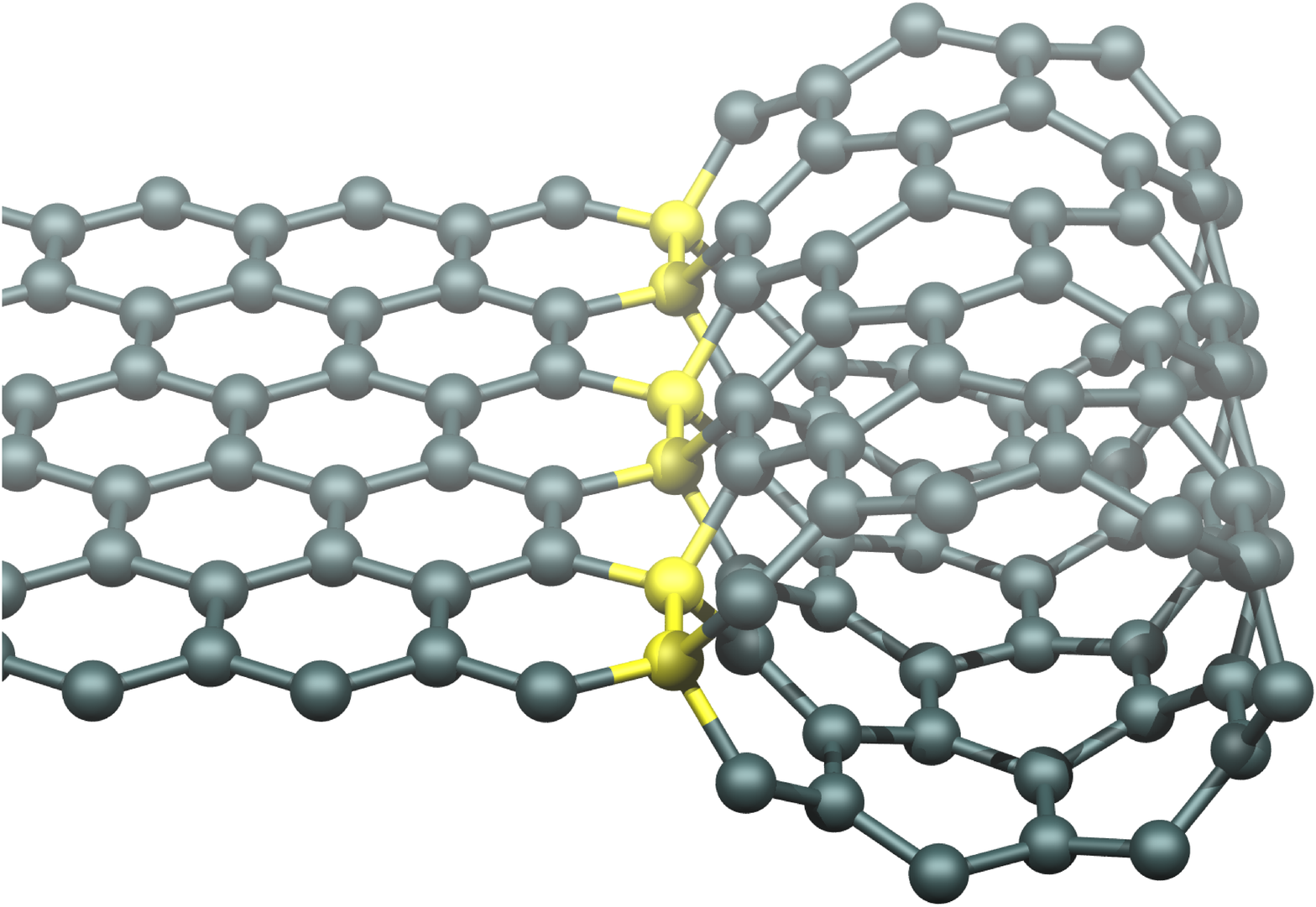}}
 }

\addtocounter{subfigure}{-3}
 \subfigure[]{
 \includegraphics[width=0.3\linewidth]{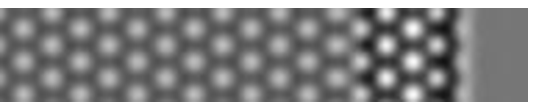}
 }
 \subfigure[]{
 \includegraphics[width=0.3\linewidth]{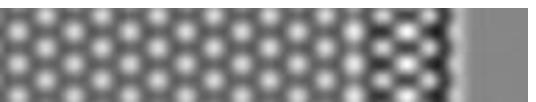}
 }
 \subfigure[]{
 \includegraphics[width=0.3\linewidth]{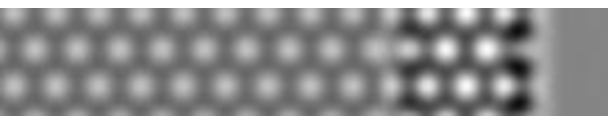}
 }

  \caption{(Color online) Upper panel. Structure of (4,4) armchair (a), (4,4)
armchair-like (b) and (8,0) zigzag (c) nanotube terminated graphene sheet.
sp$^3$-like coordinated carbon atoms are marked in yellow.
 Bottom panel. Corresponding simulated HRTEM images.
% NOTE FOR THE EDITOR: IF THERE IS SUFFICIENT SPACE TO WIDEN THIS TO FULL PAGE WIDTH
% IT WOULD BE CLEARER, OTHERWISE IS OK LIKE THIS...
}
  \label{fig}
\end{figure}

A new type of termination leading to the stabilization of the edge can be
achieved by taking an unreconstructed free edge, folding it back on itself and
bonding the edge dangling bonds to a line of basal graphene atoms.
A rolled zigzag edge can bond into the graphene plane in two configurations:
either above what would be zigzag edge atoms or Klein edge atoms. 
Rolled armchair edges can only bond to equivalent armchair edge atoms.
Drawing on nanotube nomenclature we refer to these new
edges as armchair-nanotube terminated (Fig. \ref{fig}a), armchair-like
nanotube terminated (Fig. \ref{fig}b), and zigzag-nanotube terminated (Fig.
\ref{fig}c) respectively.
In all cases the line of carbon atoms bridging the tube and graphene layer
adopts an sp$^3$-like hybridization with average bond lengths ($\sim 1.50$\AA{})
and angles ($\sim108^{\circ}$) close to those of diamond.
Locally, the structure is similar to the core of the zigzag
prismatic dislocation in AA graphite \cite{suarez2007dislocations}.
The sp$^3$-like bonding allows the
tube to localise strain, resulting in a droplet-shaped
cross-section (Fig. \ref{fig}).

Fig. \ref{formenerg} presents formation energies for different
types of free and tube terminated edges as a function of the Hamada
index of the tube.
For comparison we plot the values for free-standing armchair and zigzag
tubes. Consistent with an earlier study \cite{Sawada1992917}, free
standing small tubes with diameters below 4 \AA{} (i.e. (3,3) and
(5,0) and below) are unstable compared to a flat graphene sheet. This means
that it is thermodynamically preferable to split these tubes open, even with
unfunctionalised edges. Indeed experimentally these small radii nanotubes have
not been observed on their own but they can exist as inner tubes in large
multiwalled
nanotubes \cite{qin2000materials,*guan2008smallest}. 
We see here however that such small nanotubes are more stable when formed on
graphene ribbon
edges, through localisation of the nanotube strain along the sp$^3$-coordinated
tetrahedral bonding line, e.g. a (3,3) armchair tube and all zigzag tubes up to
(7,0).
Comparing to free zigzag, edges become more stable when rolled in tubes above
(8,8).
Rolled armchair edges have a lower energy than free edges when forming nanotubes
larger than (14,0).
For the largest presented tube terminated edges, formation energies are
lower than any
of the previously proposed free edge configurations.

If we extrapolate to larger diameter tubes we might expect that
they collapse due to Van der Waals interactions
between walls to a dog-bone cross-section \cite{chopra1995fully}.
However the droplet cross-section induced by the line of sp$^3$-like carbon
atoms naturally induces a ``local collapse'', and this pinched region extends
further as the tube diameter increases. Thus for large diameters, rolled edges
converge to a classical folded edge which then terminates some distance
from the actual edge via a line of sp$^3$-like bonds.
We note that in no case is the combination of a free tube and graphene edge more
stable than the nanotube terminated edge \cite{Chernozat-1} {\it i.e.} there will be no
thermodynamic driving force for nanotube production from rolled edges. 

We next examined the barrier to roll up the edges, determining the barrier
between the free zigzag edge and an armchair-nanotube terminated edge
as a function of the final tube size.  
For a (4,4) nanotube terminated edge, we find an NEB barrier for
ribbon rolling of about 2 eV/\AA, with a corresponding unrolling barrier
of 0.9 eV/\AA.  For a (8,8) nanotube terminated edge barriers change to 1.3
eV/{\AA} to roll versus 1.6 eV/{\AA} for derolling.
Increasing further the tube diameter
gives tube reconstructed edges more stable than free edges and this induces
a barrier
height inversion.
In particular for small tubes these barriers are quite high, due to the high
curvature that has to be induced. However these represent
maximum barriers and experimental barriers will likely
be much smaller, since our calculations assume concerted simultaneous
bonding/debonding along the entire edge length.  Similar to dislocation motion,
which proceeds via
the propagation of kinks along the dislocation line rather than a concerted
single-step motion, rolled edges will presumably roll/unroll initially at a
single point which will then propagate along the nanotube length.

Given their low formation energies compared to free zigzag edges, we can
ask the question why such tube terminated edges have not yet been reported in
the literature. One reason is that these edges require long-range
order, whereas other edge structures can vary over the order of single unit
cells. Thus we may expect such edges to be more common in well defined periodic
ribbon edges such as after splitting of large
multi-walled nanotubes.
We note also that free standing graphene can be obtained through wet etching
after epitaxial growth on metal substrate \cite{shivaraman2009free}. We expect
that during the etching process rolled edges might appear as in analogous
synthesis mechanisms used for the production of inorganic nanotubular materials
\cite{schmidt2001nanotechnology}. 
Furthermore, rolled edges could be difficult to discriminate by transmission
electron microscopy. Simulated HRTEM images (Fig. \ref{fig}) are very
similar to those of free standing edges, the primary difference being minor
variations in the image contrast.  Additionally these edges may display
characteristic nanotube modes detectable using spatially resolved resonant
Raman.

\begin{figure}
\includegraphics[width=\linewidth]{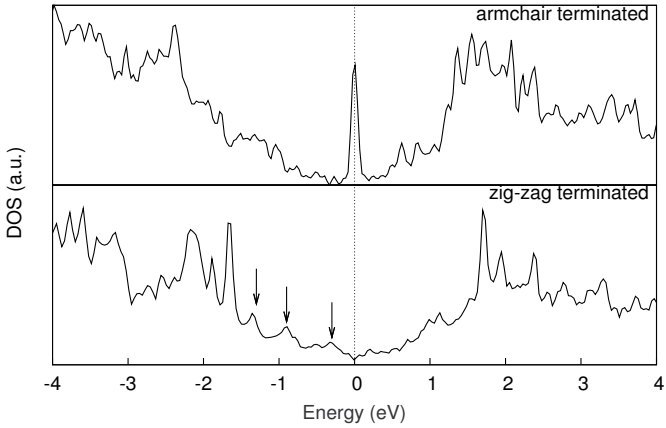}
 \caption{Calculated density of states for (a) (8,8) armchair- and (b) (8,0)
 zigzag-nanotube terminated graphene edge. The Van Hove
singularities are marked by arrows.}
\label{dos}
\end{figure}

We next examine the effect of rolled edges on the electronic properties of the
graphene.
We present in Fig. \ref{dos} the electronic density of states for the
(8,8) and (8,0)
tube
terminated edges. These show an interesting combination of the graphene and
nanotube behaviour.
In spite of different electronic character of free standing (8,8) and (8,0)
tubes, {\it i.e.} metallic and semiconducting, in both cases the composite
system has
a non-zero density of states at the Fermi level. 
For a zigzag tube termination the background density of
states around the Fermi level rises smoothly, reflecting the graphene density of
states, overlaid on which there is a series of Van Hove singularities
characteristic
of a nanotube. 
For the armchair-tube terminated edge, {\it i.e.} a rolled zigzag graphene edge,
there is a sharp peak at the Fermi level similar to that seen for flat
unterminated
zigzag edges \cite{Pekka, kunstmann2011stability}.  Such Fermi level peaks can
lead to magnetic
instability, and indeed our spin unrestricted calculations shows a ferromagnetic
configuration to be the lowest energy state (slightly lowering
of the edge energy of the system by only 0.02 eV/\AA), although this must be
treated with caution given the use of the local density
approximation.

\begin{figure}[htb]
\centering
 \subfigure[]{
{\includegraphics[width=0.55\linewidth]{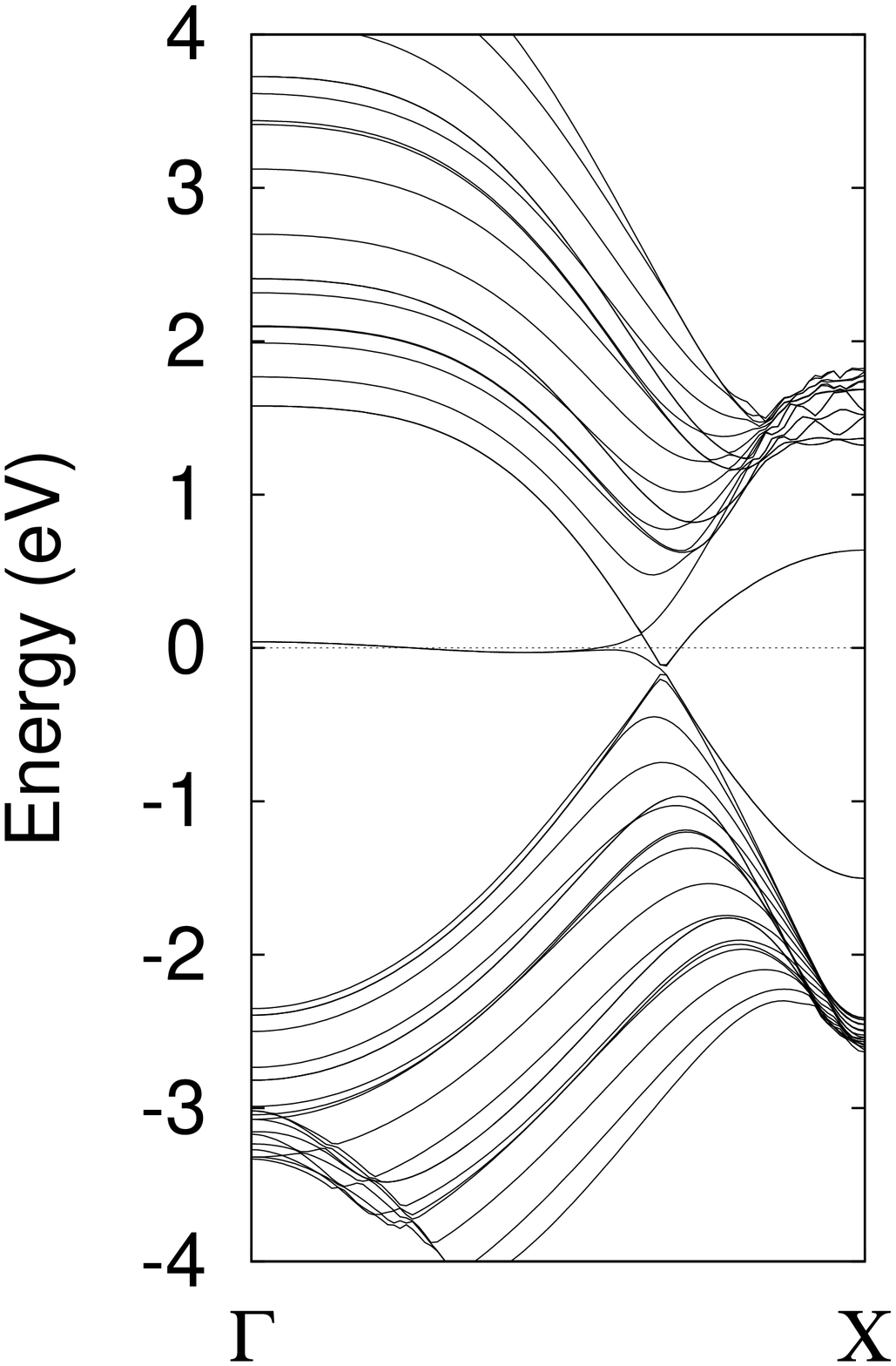}}
 }
 \subfigure[]{
{
\includegraphics[angle=0,width=0.35\linewidth]{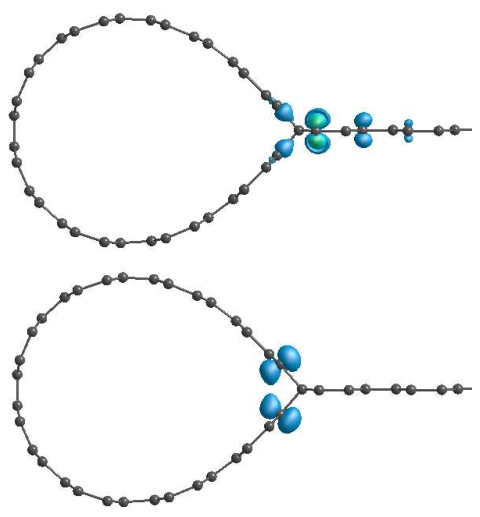}}
 }
  \caption{(Color online) (a) Band structure for (8,8) armchair-nanotube
terminated edges,
  and (b) distribution of highest occupied state at the $\Gamma$-point (up) and
$X$-point (down).
  }
  \label{bandst}
\end{figure}

The peak consists of two degenerate states which can be seen in the associated
band structure (Fig. \ref{bandst}a) as involving a mixing of several bands. 
Plotting these
states at the $\Gamma$ and $X$ point (Fig. \ref{bandst}b) shows that the
Fermi level spike is localised mainly on the row of graphene atoms next to the
sp$^3$-carbon atoms,
mirroring the edge state seen in flat zigzag terminated graphene
\cite{enoki2007electronic, *kunstmann2011stability, *wassmann2008structure},
whereas at the $X$-point
the highest occupied state comes from dispersive edge states located in the
nanotube segment
along the junction.  This convergence of three zigzag edges at the line of
sp$^3$-bonded
carbon atoms suggests possible interesting magnetic behaviour under an applied
field.

We note that the zigzag metallic edge state is preserved in this configuration,
and unlike the unterminated simple zigzag edge it will also be partially
protected from environmental attack, since all neighbouring atoms are fully
coordinated. 
Notably we might expect it to be stable in air. These states also suggest
intriguing transport behaviour, with possible conduction channels both along
the edge of the graphene and in the edge states in the nanotube segment.  

In summary we have shown that by rolling an unterminated graphene edge it is possible to create
nanotube-terminated edges where the sheet edge rebonds back into the graphene
plane.  We determine the critical tube size and formation barriers and compare
with density functional simulations of other edge terminations including a
new reconstructed Klein edge.
We find that the proposed tube terminated edges are more stable than any other
non-functionalized edge structure, due to the replacement of dangling bonds with
sp$^3$-like hybridised carbon atoms.
Rolled zigzag edges serve as metallic conduction channels,
separated from the neighbouring bulk graphene by a chain of insulating
sp$^3$-carbon atoms, and introduce Van Hove singularities into the graphene
density of states. 
They may provide a way to stabilise and protect from chemical attack
the disperse Fermi level state seen along metallic zigzag edges.
Similar edge rolling effects might also
appear in other
layered materials \cite{coleman2011two} such as boron nitride monolayers where
orbital rehybridization can occur.
\acknowledgments
This work has been carried out within the NANOSIM-GRAPHENE project nANR-09-
NANO-016-01 funded by the French National Agency (ANR) in the frame of its 2009
programme in Nanosciences, Nanotechnologies and Nanosystems (P3N2009). 

\newpage

\appendix
\section{\large{Supplementary materials.}}

\paragraph{Computational Details}

Structures were geometrically optimized using a ($1 \times1 \times12$) shifted
k-point grid generated using the Monkhorst-Pack formalism.  Hartwigsen,
Goedecker and Hutter relativistic pseudopotentials were used\cite{Hgh-98}.
Atom-centered Gaussian basis functions are used to construct the many-electron
wave function with angular momenta up to l=2, i.e. 22 independent functions per
C atom. 
We confirmed that all structures were energetically converged with respect to 
ribbon width by extending some of the calculations by another hexagon row,
finding formation energy changes corresponding to the addition of bulk	
graphene atoms.
Electronic level occupation was obtained using a Fermi occupation function with
kT = 0.04 eV. 
Density of states calculations used a denser ($1 \times1 \times100$) k-point
grid, with a 0.03 eV Gaussian broadening applied.

\begin{figure*}[H]
\begin{tabular*}{0.7\textwidth}{@{\extracolsep{\fill}}ccc}
{\includegraphics[width=0.14\linewidth]{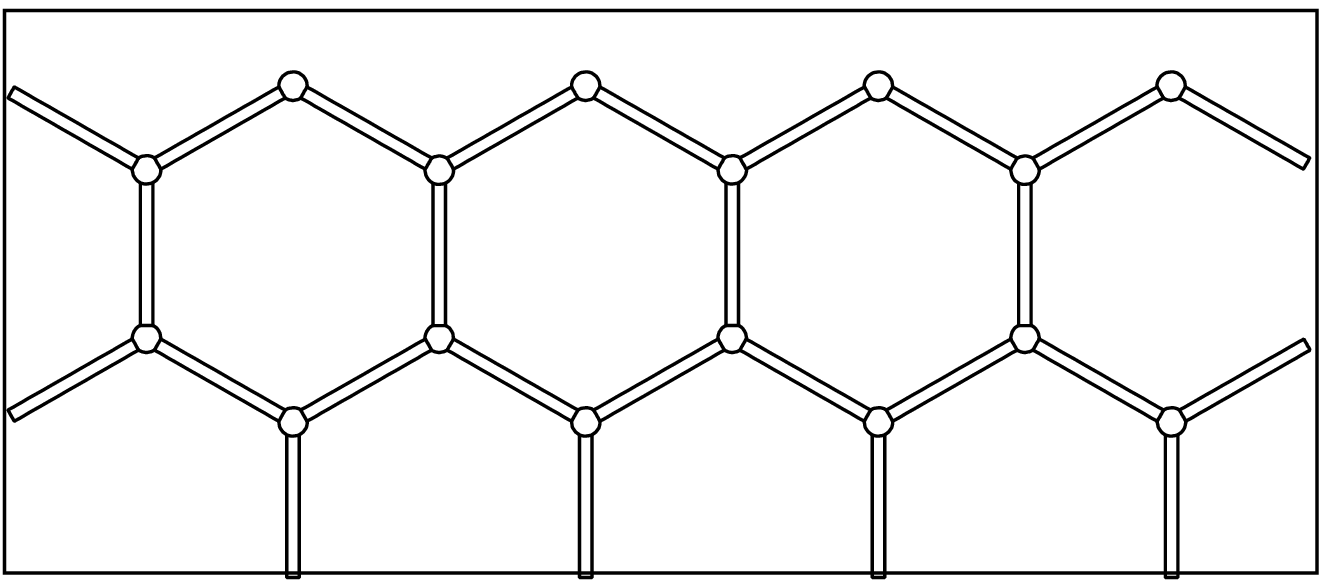}} & 
{\includegraphics[width=0.14\linewidth]{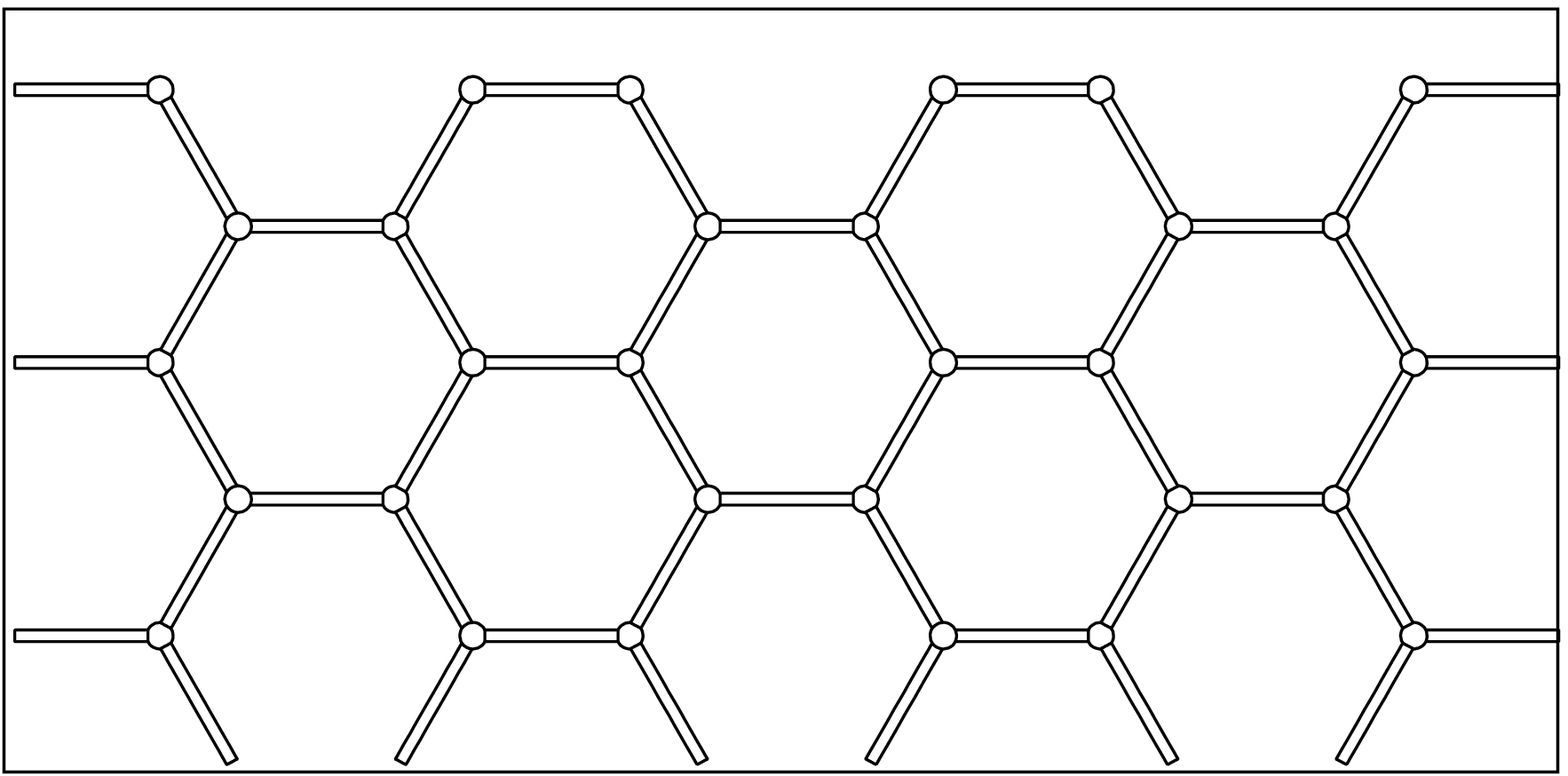}}&
{\includegraphics[width=0.14\linewidth]{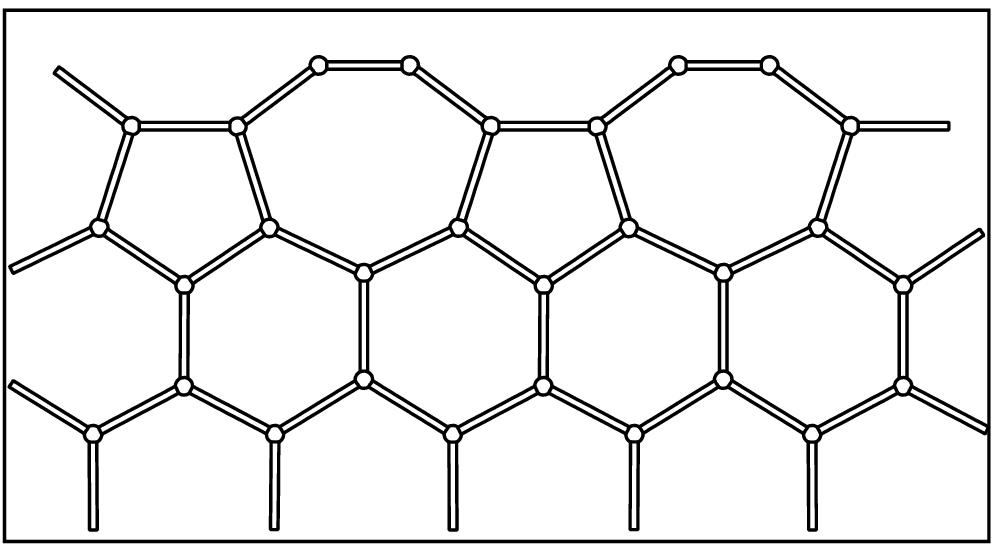}}\\

{\includegraphics[width=0.14\linewidth]{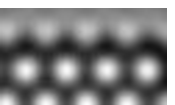}} &
{\includegraphics[width=0.14\linewidth]{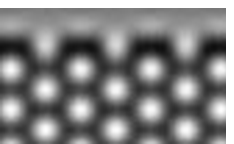}} &
{\includegraphics[width=0.14\linewidth]{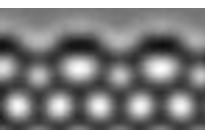}} \\

Zigzag & Armchair  & 5-7 reconstructed zigzag \\
 1.34 eV/\AA & 1.10 eV/\AA & 1.09 eV/\AA \\

{\includegraphics[width=0.14\linewidth]{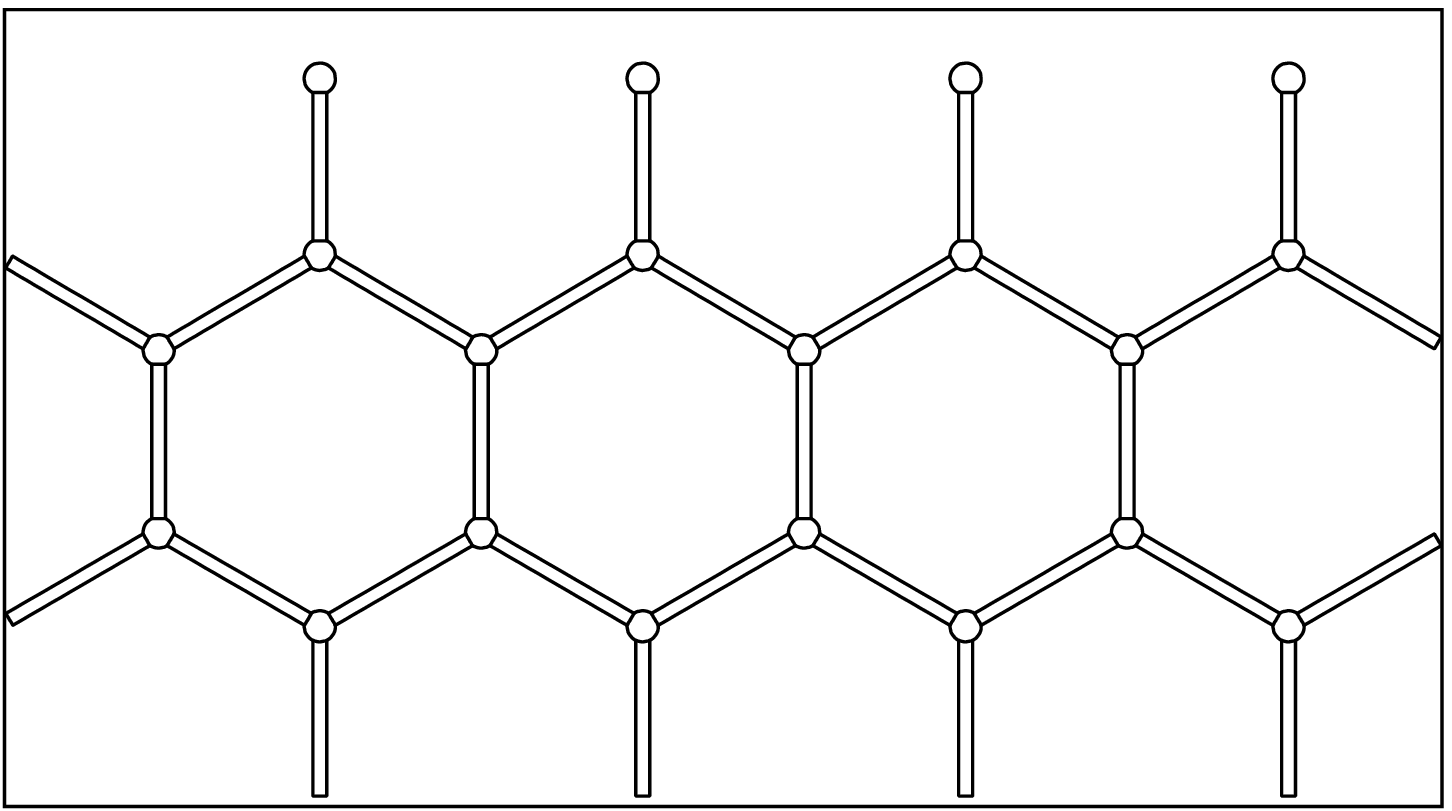}}&
{\includegraphics[width=0.14\linewidth]{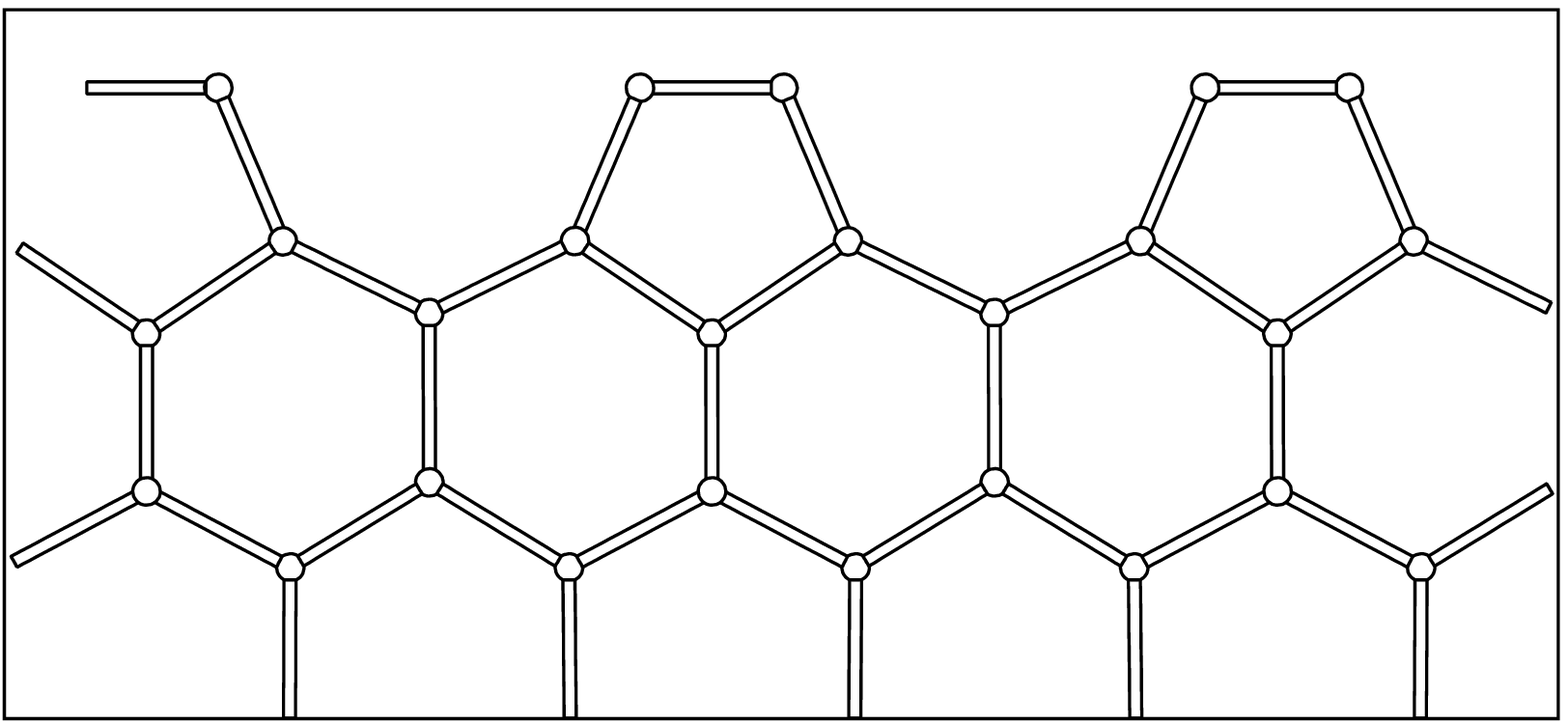}}&
{\includegraphics[width=0.14\linewidth]{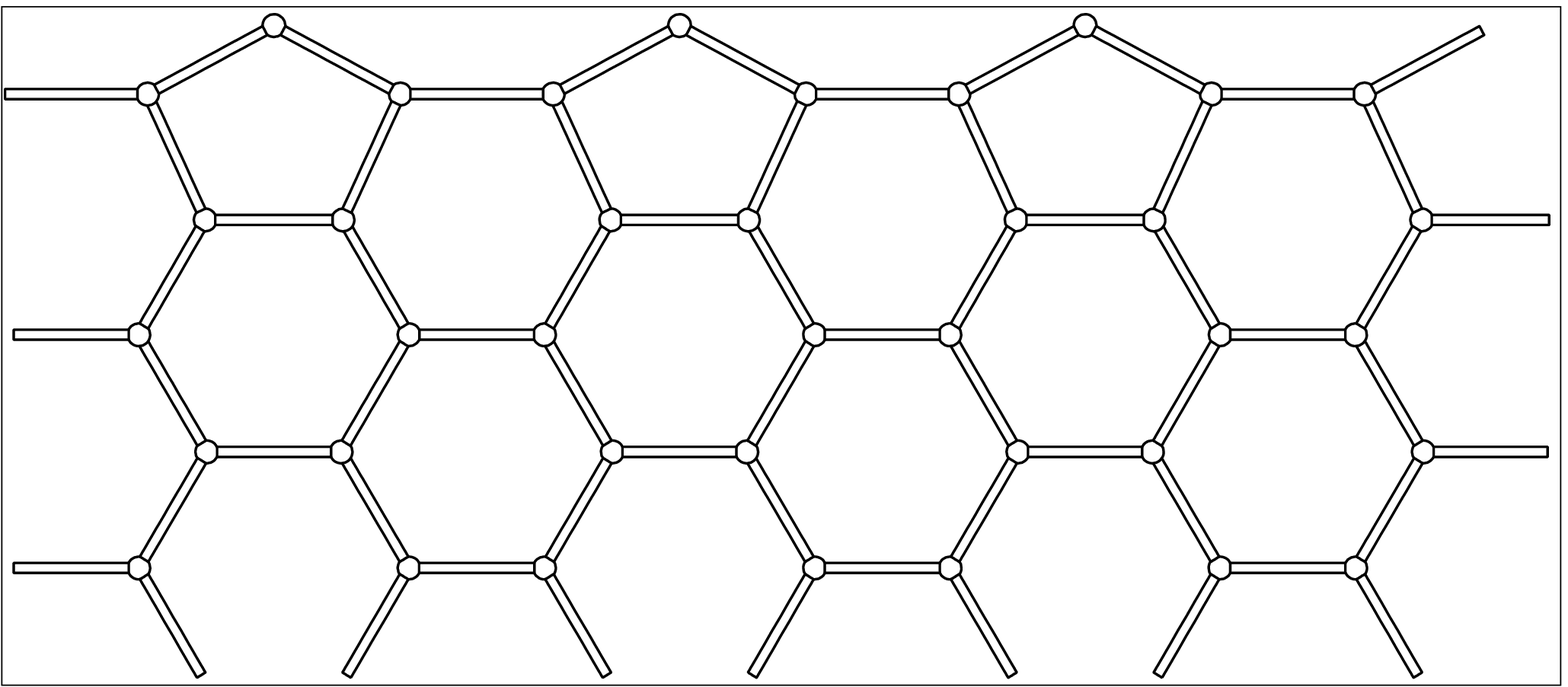}}\\

{\includegraphics[width=0.14\linewidth]{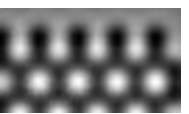}} & 
{\includegraphics[width=0.14\linewidth]{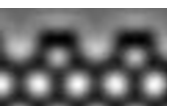}} &
{\includegraphics[width=0.14\linewidth]{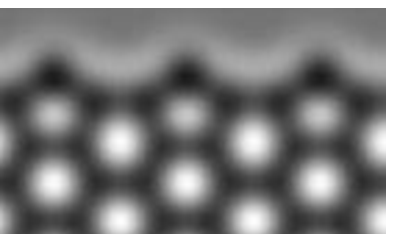}}\\
 Klein
edge  & Reconstructed Klein edge & Reconstructed armchair \\
 2.21 eV/\AA &  1.50 eV/\AA & 1.63 eV/\AA\\
\end{tabular*}
\caption{Calculated formation energies  for different edge
structures and corresponding simulated HRTEM images.}
\label{Tab}
\end{figure*}

\begin{figure}[tbp]
 \includegraphics[width=0.3\textwidth]{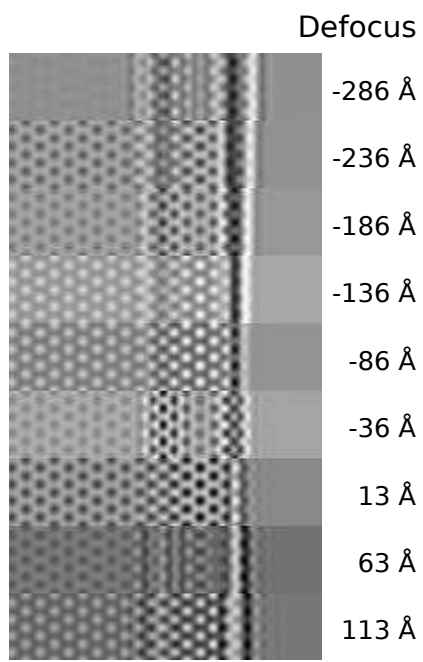}
\caption{HRTEM simulated focal series images for (8,8) armchair tube terminated
edge.}
\label{focal-series}
\end{figure}

\paragraph{HRTEM image simulations}
High resolution transmission electron microscopy (HRTEM) images were
simulated using a full dynamical calculation by multislice method as
implemented in the SimulaTEM program \cite{SimTEM}. 
Formation energies and HREM image simulation for different edge types are
presented in Fig. \ref{Tab}.
A focal series for a (8,8)
armchair rolled tube is presented in Fig. \ref{focal-series}.

\end{document}